\documentclass[aps,superscriptaddress,letterpaper,amsmath,twocolumn]{revtex4}
\usepackage{graphicx}
\usepackage{dcolumn}
\usepackage{bm}
\usepackage{amssymb}
\usepackage{array}
\usepackage{longtable}
\usepackage{CJK}
\usepackage{verbatim}
\usepackage{color}

\begin{document}


\title{Mean first-passage time for maximal-entropy random walks in complex networks}

\author{Yuan Lin}

\author{Zhongzhi Zhang}
\email[Email of the corresponding author: ]{zhangzz@fudan.edu.cn}

\affiliation {School of Computer Science, Fudan University,
Shanghai 200433, China}

\affiliation {Shanghai Key Lab of Intelligent Information
Processing, Fudan University, Shanghai 200433, China}

\begin{abstract}
We perform an in-depth study for mean first-passage time (MFPT)---a primary quantity for random walks with numerous applications---of maximal-entropy random walks (MERW) performed in complex networks. For MERW in a general network, we derive an explicit expression of MFPT in terms of the eigenvalues and eigenvectors of the adjacency matrix associated with the network. For MERW in uncorrelated networks, we also provide a theoretical formula of MFPT at the mean-field level, based on which we further evaluate the dominant scalings of MFPT to different targets for MERW in uncorrelated scale-free networks, and compare the results with those corresponding to traditional unbiased random walks (TURW). We show that the MFPT to a hub node is much lower for MERW than for TURW. However, when the destination is a node with the least degree or a uniformly chosen node, the MFPT is higher for MERW than for TURW. Since MFPT to a uniformly chosen node measures real efficiency of search in networks, our work provides insight into general searching process in complex networks.
\end{abstract}




\maketitle



Random walks in complex networks have been heavily studied in the past years~\cite{NoRi04,MeKl04,BuCa05} due to their wide range of applications in science and engineering~\cite{We1994}. Recently, continuously increasing efforts have been devoted to maximal-entropy random walks (MERW)~\cite{BuDuLuWa09,BuDuLuWa10,SiGoLaNiLa11,Oc12B,FrGa14}, also called Ruelle-Bowens random walks~\cite{Ru04,Pa64}, where all walking trajectories from given starting and ending points of a given length are equiprobable. In this sense, MERW is the most random of random walks, which  maximizes the entropy rate~\cite{CoTh91,GoLa08} and is in striking contrast with the traditional unbiased random walks (TURW) and other biased random walks. 
The unique diffusion process of MERW leads to several unusual phenomena, such as localization of stationary distribution~\cite{BuDuLuWa09} and fast relaxation~\cite{OcBu12}.

As a powerful tool, MERW has been fruitfully applied in various fields. For instance, the localization phenomenon of stationary distribution for MERW  makes it a good measure of  centrality that is more discriminating than some classic centrality measures, e.g. PageRank, in the sense that it can distinguish evidently those nodes that PageRank regards as almost equally important~\cite{DeLi11,Oc12}. Furthermore, since MERW incorporates both network structure and eigenvector centrality of nodes, it was also applied to establish a new algorithm of link prediction, which outperforms various supervised and unsupervised techniques of link prediction, on most test databases~\cite{LiYuLi11}. In addition, MERW has also found applications in optimal sampling algorithm~\cite{He84}, demographic stability of population~\cite{DeGuOc04}, community detection~\cite{OcBu13}.

A fundamental quantity related to random walks is first-passage time (FPT)~\cite{Re01,CoBeMo05,CoBe07,CoBeTeVoKl07}, which is the expected time required for a random walker starting from a source point to a given target point. The mean first-passage time (MFPT) is defined as the average of FPTs over all source nodes in the network, which is a useful tool to analyze the behavior of random walks. The importance of MFPT originates from the essential role played by first encounter features appearing in various real situations, such as lighting harvesting~\cite{BaKlKo97,BaKl98,BeHoKo03} and target search~\cite{JaBl01,BeLoMoVo11}. The MFPT can also serve as a significant indicator measuring node importance~\cite{WhSm03} and efficiency of trapping process~\cite{CaAd08}. It is thus of utmost importance to study MFPT for different random-walk processes. Thus far, MFPT has been intensively studied for TURW~\cite{Ag08,KaBa02PRE,ZhQiZhXiGu09,AgBu09,TeBeVo09,AgBuMa10,MeAgBeVo12,LiJuZh12,HwLeKa12} and some biased random walks~\cite{WuZh13,YaZh13,FrFr09,BoNiLa14}, while related research about MFPT for MERW is still much less, although the particular diffusion process of MERW is suggested to significantly affect the leading behavior of MFPT.

In this paper, we propose a theoretical framework for MERW in complex networks and perform an in-depth study on the MFPT for MERW to a given target. We derive an explicit expression of FPT for MERW from one node to another in any connected network in terms of the eigenvalues and eigenvectors of adjacency matrix for the network. Based on the obtained representation for FPT, we further deduce an exact formula for MFPT to an arbitrary target node. Moreover, for uncorrelated networks, we also provide an analytical expression of MFPT for MERW at the mean-field level, using which we obtain the leading scalings of MFPT for uncorrelated scale-free networks with various degree exponent $\gamma$, and show how the MFPT scales with the network size.

For MERW in uncorrelated scale-free networks, we study the MFPT for three representative cases with the target node being located at a hub node, a node with the smallest degree, and a node uniformly chosen from the system, respectively. For all the three cases, we derive analytically the leading scalings for MFPT, all of which depend on the degree exponent $\gamma$ that characterizes the heterogeneity extent of scale-free networks. Our results indicate that for the last two cases that the target is placed at a smallest degree node or a uniformly selected node, the leading scalings resemble each other, but both scalings are considerably largerx than that corresponding to the first case when a hub is the target.

We also compare the obtained results of MFPT for uncorrelated scale-free networks with those corresponding to TURW. We show that when the target is fixed at a hub node, the MFPT for MERW is much less than that for TURW. On the contrary, when the target is placed at a smallest degree node or a randomly chosen node, the MFPT for MERW is larger than that associated with TURW. Therefore, in comparison with TURW, the special diffusion process of MERW has a stronger influence on the efficiency for searching a target in heterogeneous networks, making the process considerably more efficient for finding hub node but less efficient for locating a node with small degree or a randomly chosen node.


\section*{\large{Results}}


\textbf{Explicit expressions of MFPT for MERW.} Throughout the paper, the random walk processes considered are defined in a connected undirected graph $G$ with $N$ nodes and $E$ edges. The connectivity of nodes is described by the adjacency matrix ${\bf A}$, in which the entry $a_{ij}=1$ if nodes $i$ and $j$ are adjacent, and $a_{ij}=0$ otherwise. Then, the degree of a node $i$ is $k_i=\sum_{j=1}^{N}a_{ij}$. Let $\lambda_1, \lambda_2, \cdots, \lambda_N$ be the $N$ eigenvalues of $\bf A$, rearranged as $\lambda_1>\lambda_2\geq\cdots\geq\lambda_N$, and let $\mu_1, \mu_2, \cdots, \mu_N$ be their corresponding mutually orthogonal eigenvectors of unit length, where $\mu_i=(\mu_{i1}, \mu_{i2}, \cdots, \mu_{iN})^{\top}$. Then, matrix $\bf A$ can be decomposed as the following spectral form:
\begin{equation}\label{A3}
{\bf A}={\bf U}{\rm diag}[\lambda_1, \lambda_2, \cdots, \lambda_N]{\bf U}^{\top},
\end{equation}
where ${\bf U}=(\mu_1, \mu_2, \cdots, \mu_N)$ is an orthogonal matrix, obeying
\begin{equation}\label{A4}
{\bf U}{\bf U}^{\top}={\bf U}^{\top}{\bf U}={\bf I},
\end{equation}
where $\bf I$ is the identity matrix.

Using above notations, we can introduce the MERW that is characterized by a unique stochastic matrix {\bf P}, the $ij$th entry of which is given by
\begin{equation}\label{A5}
p_{ij}=\frac{a_{ij}}{\lambda_1}\frac{\mu_{1j}}{\mu_{1i}}\,,
\end{equation}
where $\lambda_1$ is the principal eigenvalue of matrix ${\bf A}$, and $\mu_{1i}$ is the $i$th entry of the principal eigenvector $\mu_{1}$ corresponding to $\lambda_1$.  This guarantees that MERW maximizes the entropy of a set of trajectories with a given length and end-points and leads to the maximal entropy rate of such processes~\cite{BuDuLuWa09}. The stationary distribution of MERW is~\cite{Pa64}
\begin{equation}\label{A6}
\pi=(\pi_1, \pi_2, \cdots, \pi_N)^{\top}=(\mu_{11}^2, \mu_{12}^2, \cdots, \mu_{1N}^2)^{\top}\,.
\end{equation}
The MERW is biased, which is different from TURW, where the transition probability $p_{ij}=a_{ij}/k_i$ from a node $i$ to one of its neighboring nodes $j$ is identical.

The main quantity we are interested in the paper is MFPT. Notice that MERW in an arbitrary connected binary network can be represented as generic random walk in a corresponding weighted network~\cite{LaSiDeEvBaLa11}. The ${ij}$th element of the generalized adjacency matrix (weight matrix) ${\bf W}$ for the weighted network is defined by $w_{ij}=a_{ij}\mu_{1i}\mu_{1j}$, which specifies the weight of the edge connecting nodes $i$ and $j$. In this weighted network, the strength~\cite{BaBaVe04} of a node $i$ is given by $s_i=\sum_{j=1}^{N}w_{ij}=\lambda_1\mu_{1i}^2$, and the total strength of all nodes is $s=\sum_{i=1}^{N}s_i=\lambda_1$. For generic random walks in this weighted network, the transition probability is defined as
\begin{equation}\label{A7}
p_{ij}=\frac{w_{ij}}{s_i}=\frac{a_{ij}\mu_{1i}\mu_{1j}}{\lambda_1\mu_{1i}^2}=\frac{a_{ij}\mu_{1j}}{\lambda_1\mu_{1i}},
\end{equation}
which is equal to transition probability, given by equation~(\ref{A5}), for MERW in the original graph. The equivalence between the two random walks allows to determine the MFPT for MERW in a graph by evaluating the corresponding quantity for generic random walks in a related weighted network.



For MERW in a network, let $T_{ij}$ denote the FPT from node $i$ to node $j$. Without loss of generality, let $j$ be the target node, and let $T_j$ be the MFPT to node $j$. Then, by definition, the MFPT $T_j$ is given by
\begin{equation}\label{B0}
T_j=\frac{1}{N-1}\sum_{i=1}^{N}T_{ij}\,.
\end{equation}
Based on the equivalence between MERW and corresponding generic random walks, we derive an exact expression (see Methods) for $T_{ij}$ in terms of the eigenvalues and eigenvectors for adjacency matrix $\bf A$:
\begin{equation}\label{B7x}
T_{ij}=\frac{1}{\mu_{1j}^2}\sum_{k=2}^N\frac{\lambda_1}{\lambda_1-\lambda_k}\left(\mu_{kj}^2-\mu_{ki}\mu_{kj}\frac{\mu_{1j}}{\mu_{1i}}\right)\,.
\end{equation}
Plugging equation~(\ref{B7x}) into equation~(\ref{B0}), we arrive at an expression of MFPT $T_j$ for MERW in a general graph with the deep trap fixed at an arbitrary node $j$, given by
\begin{equation}\label{B8}
T_j=\frac{1}{\mu_{1j}^2(N-1)}\sum_{k=2}^{N}\frac{\lambda_1}{\lambda_1-\lambda_k}\left(N\,\mu_{kj}^2-\mu_{kj}\mu_{1j}\sum_{i=1}^{N}\frac{\mu_{ki}}{\mu_{1i}}\right)\,.
\end{equation}

Equation~(\ref{B8}) provides a universal formula of MFPT to any node for MERW in an arbitrary network. Although it involves computing eigenvalues and eigenvectors of adjacency matrix, which puts heavy demands on time and computation resources for large networks, it can be utilized to check the results for MFPT obtained by other approaches, at least for
small networks. Besides, equation~(\ref{B8}) can also be used to compute the exact average $\langle T \rangle$ of $T_j$ over all $N$ targets:
\begin{small}
\begin{eqnarray}\label{B9}
& &\langle T \rangle=\frac{1}{N}\sum_{j=1}^{N} \,T_j\nonumber \\
&=&\frac{1}{N(N-1)}\sum_{j=1}^{N}\frac{1}{\mu_{1j}^2}\sum_{k=2}^{N}\frac{\lambda_1}{\lambda_1-\lambda_k}\left(N\,\mu_{kj}^2-\mu_{kj}\mu_{1j}\sum_{i=1}^{N}\frac{\mu_{ki}}{\mu_{1i}}\right),
\end{eqnarray}
\end{small}
which is exactly the MFPT when the target is uniformly distributed.

A drawback for equation~(\ref{B8}) is that by using this spectral technique it seems very difficult, even impossible, to obtain the leading behavior of MFPT $T_j$ characterizing the random-walk dynamic process. Thus, it is important to seek alternative techniques of evaluating MFPT $T_j$ even for particular
networks, which are not computationally demanding but are valid to estimate the scaling of MFPT. Fortunately, for uncorrelated networks, we can derive an expression of MFPT for MERW at the mean-field level, as well as its dominant scaling for scale-free networks. The details will be given below.

\textbf{Theoretical prediction of MFPT for MERW in uncorrelated networks.} We now consider the MFPT for MERW in uncorrelated networks, where the degree-degree correlations between adjacent nodes are absent. In a recent work~\cite{LiZh13}, we have shown that, for generic random walks in uncorrelated weighted networks, the MFPT to node $j$ can be represented in terms of the strengths of the nodes as
\begin{equation}\label{C1}
T_j=\frac{s}{s_j}\,.
\end{equation}
Plugging $s_j=\lambda_1\mu_{1j}^2$ and $s=\lambda_1$ into equation~(\ref{C1}) gives the MFPT to node $j$ for MERW:
\begin{equation}\label{C2}
T_j=\frac{\lambda_1}{\lambda_1\mu_{1j}^2}=\frac{1}{\mu_{1j}^2}\,.
\end{equation}
Thus, in order to obtain $T_j$, it is sufficient to determine $\mu_{1j}$. Although the evaluation of eigenvectors of a general matrix is very hard, for uncorrelated networks we can approximate $\mu_{1j}^2$ at the mean-field level (see Methods) as follows:
\begin{equation}\label{Z1}
\mu_{1j}^2\approx \frac{k_j^2}{\sum_{i=1}^{N}k_i^2}\,.
\end{equation}
Substituting equation~(\ref{Z1}) into equation~(\ref{C2}), we reach a theoretical approximation of MFPT for MERW to node $j$:
\begin{equation}\label{C11}
T_j=\frac{\sum_{i=1}^N k_i^2}{k_j^2}\,.
\end{equation}

\begin{figure}
\begin{center}
\includegraphics[width=1.0 \linewidth,trim=50 50 0 45]{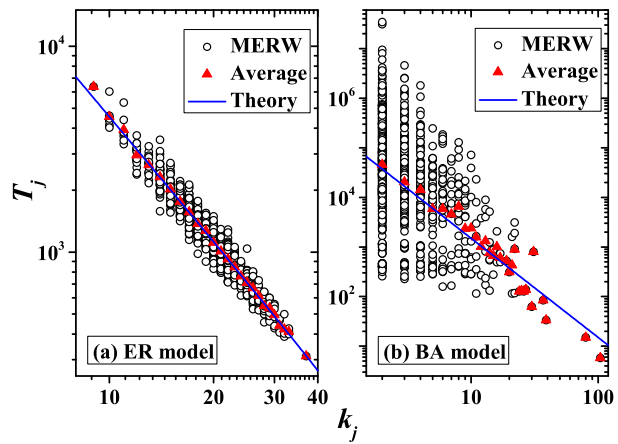}
\end{center}
\caption[kurzform]{\textbf{MFPT to a given node for MERW in ER network (a) and BA network (b).} Black circles represent the numerical results obtained by equation~(\ref{B8}), and each red triangle stands for the average of numerical values for $T_j$ over all nodes having the same degree $k_j$. Straight lines are the theoretical approximation generated according to equation~(\ref{C11}). 
}\label{MERW-Theory}
\end{figure}

In Fig.~\ref{MERW-Theory}, we report both the exact numerical results and theoretical approximate results of MFPT for MERW taking place in Erd\"os-R\'enyi (ER) network~\cite{ErRe60} and Barab\'asi-Albert (BA) network~\cite{BaAl99}, with both results being generated by equations~(\ref{B8}) and~(\ref{C11}), respectively. Figure~\ref{MERW-Theory} shows that the theoretical predictions agree well with the numerical results. From Fig.~\ref{MERW-Theory}, we can also find that for nodes sharing identical degree, MFPT for MERW distributes in a broader range in BA network than in ER network. This difference lies in the structure of the networks. Since BA networks is heterogeneous, the component of leading eigenvector localizes at hub nodes and their neighbors~\cite{BuDuLuWa09}.  For nodes having the same degree but different neighbors, their MFPT differ widely. For example, for two leaf nodes in treelike BA network that are linked to a hub node and a small-degree node farther from the hub, respectively, the MFPT to the leaf connected to a hub is much less than the MFPT to the other leaf. While for ER network, it is almost homogeneous, so the disparity for MFPT to different nodes with identical degree is relatively indiscernible.

In order to better understand the behavior of MFPT in inhomogeneous networks,  in the sequel, grounded on the theoretical approximation in equation~(\ref{C11}), we will analytically evaluate the leading scaling of MFPT for MERW in uncorrelated scale-free networks, aiming to unveil the effects of target location on the MFPT for MERW, as well as the difference between MERW and TURW in terms of the MFPT.




\textbf{Leading scalings of MFPT for MERW in uncorrelated scale-free networks.} Extensive empirical studies~\cite{Ne03} have shown that most real-world networks exhibit the striking scale-free property~\cite{BaAl99}, characterized by a power-law degree distribution $P(k)\sim k^{-\gamma}$ with $\gamma>2$. In this section, we will study the leading scalings of MFPT for MERW in uncorrelated scale-free networks. We will examine the dominant  scalings of MFPT for three representative target problems, with the target being a hub with the highest degree, a node with the lowest degree, or a node selected uniformly. Our goals are twofold. One is to uncover of the influence of target location or degree on the behavior of MFPT. The other is to find the scaling difference of MFPT between MERW and TURW.



\emph{Scaling of MFPT to a hub node.} Let $k_{\rm max}$ denote the degree of a hub node, and  $T_{\rm H}$ the MFPT to this hub node. Then, by equation~(\ref{C11}),
\begin{equation}\label{E1}
T_{\rm H}=\frac{\sum_{i=1}^N k_i^2}{k_{\rm max}^2}\,.
\end{equation}
The numerator in equation~(\ref{E1}) can be evaluated as
\begin{eqnarray}\label{E2}
\sum_{i=1}^N k_i^2&\approx& \int_{k_{\rm min}}^{k_{\rm max}}N P(k)k^2 dk\nonumber\\
&\sim&\left\{\begin{aligned}
&N k_{\rm max}^{3-\gamma},\quad 2<\gamma<3, \\
&N \ln k_{\rm max},\quad \gamma=3, \\
&N, \quad \gamma>3.\\
\end{aligned}
\right.
\end{eqnarray}
Note that in a scale-free network with $N$ nodes and power-law degree distribution $P(k)\sim k^{-\gamma}$, the largest connectivity $k_{\rm max}$ can be estimated as~\cite{CoErAvHa00}
\begin{equation}\label{E3}
k_{\rm max}\sim N^{1/(\gamma-1)}\,.
\end{equation}
Combining equations~(\ref{E1}-\ref{E3}), we can obtain the leading scaling of $T_{\rm H}$:
\begin{eqnarray}\label{E4}
T_{\rm H}\sim\left\{\begin{aligned}
&N^{0}, \quad 2<\gamma<3, \\
&\ln N, \quad \gamma=3, \\
&N^{(\gamma-3)/(\gamma-1)}, \quad \gamma>3. \\
\end{aligned}
\right.
\end{eqnarray}
Thus, the extent of inhomogeneity, characterized by the degree exponent $\gamma$, of scale-free networks strongly affects on the MFPT $T_{\rm H}$ to a hub node for MERW. For $2<\gamma<3$, $T_{\rm H}$ is approximately equal to a constant;  for $\gamma=3$, $T_{\rm H}$ grows logarithmically with the network size $N$; while for $\gamma> 3$, $T_{\rm H}$ grows sublinearly with $N$.

\begin{figure}
\begin{center}
\includegraphics[width=1.1 \linewidth,trim=80 80 0 50]{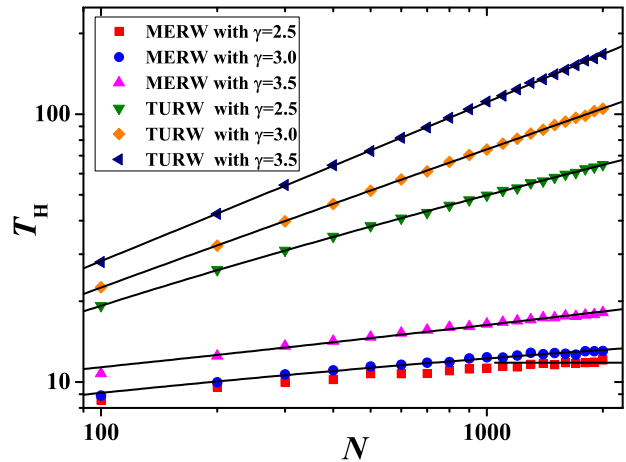}
\end{center}
\caption[kurzform]{\textbf{MFPT to a hub node as a function of the network size $N$.} 
The filled symbols are the data of numerical results. The lines correspond to theoretical predictions provided by equations~(\ref{E4}) or (\ref{E7}).
}\label{T-H}
\end{figure}

We have checked our approximate results for $T_{\rm H}$ against numerical values obtained according to equation~(\ref{B8}) for uncorrelated scale-free networks with various values of $\gamma$, namely, $\gamma=2.5$, $\gamma=3$ and $\gamma=3.5$. The considered network with $\gamma=3$ is the BA model; while the networks with $\gamma=2.5$ and $\gamma=3.5$ are generalizations of the BA model~\cite{DoMeSa00}. The comparison for theoretical and numerical results is shown in Fig.~\ref{T-H}, which indicates that for different values of $\gamma$ and network size $N$, the analytical predictions from equation~(\ref{E4}) agree with those numerical results given by equation~(\ref{B8}). It should be mentioned that for $\gamma=2.5$, the prediction is only valid for large networks, since in the process generating scale-free networks with $2<\gamma<3$, multiple and self-connections are forbidden, which could introduce degree correlations in resultant networks~\cite{MaSn02,PaNe03}, leading to a deviation of numerical values from theoretical prediction.


We next show that the behavior of MFPT for MERW in scale-free networks is quite different from that for TURW in the same networks. Similar to MFPT in weighted networks, the MFPT to a hub node for TURW in uncorrelated scale-free networks can be estimated by
\begin{equation}\label{E5}
T_{\rm H}=\frac{\sum_{i=1}^N k_i}{k_{\rm max}}\,,
\end{equation}
where the numerator $\sum_{i=1}^N k_i$ can be approximated as
\begin{equation}\label{E6}
\sum_{i=1}^N k_i\approx\int_{k_{\rm min}}^{k_{\rm max}}N P(k) k dk\sim N.
\end{equation}
Considering equation~(\ref{E3}), the leading scaling of $T_{\rm H}$ for TURW is
\begin{equation}\label{E7}
T_{\rm H}\sim N^{(\gamma-2)/(\gamma-1)}\,,
\end{equation}
which scales sublinearly with the network size $N$ but decreases with $\gamma$, a result  consistent with that previously obtained~\cite{KiCaHaAr08} using a different approach.
In Fig.~\ref{T-H}, we also plot the approximation for $T_{\rm H}$ in equation~(\ref{E7}) against their corresponding numerical values for TURW generated by the method in~\cite{LiJuZh12}, both of which agree well with each other.

Equations~(\ref{E4}) and (\ref{E7}) show that the MFPT to a hub for MERW and TURW in uncorrelated scale-free networks display rich but distinct behavior. For both MERW and TURW, the MFPT depends on the exponent $\gamma$: lower $\gamma$ corresponds to smaller MFPT. Moreover, in the whole range of $\gamma$, the MFPT for MERW is smaller than its counterpart for TURW.
The root of the difference of MFPT between TURW and MERW is attributed to their local transition probabilities. For TURW, the transition probability $p_{ij}$ from a node $i$ to a neighboring node $j$ is identical, while for MERW, the transition probability is $p_{ij}=\mu_{1j}/(\lambda_1\mu_{1i})\approx k_j/(\lambda_1 k_i)$, proportional to the degree of the neighboring node $j$. Thus, a walker visits a hub node more quickly in MERW than in
TURW.



\begin{figure}
\begin{center}
\includegraphics[width=1.1 \linewidth,trim=50 50 0 50]{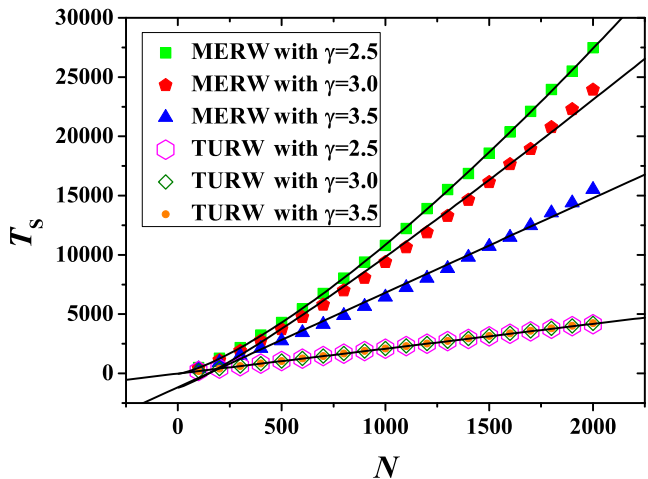}
\end{center}
\caption[kurzform]{\textbf{MFPT to the node with the least links.} 
The filled symbols stand for the numerical data, each being an average of MFPT over all nodes having the smallest degree; while the lines refer to the theoretical approximations provided by equations (\ref{X2}) or (\ref{F2}).
}\label{T-S}
\end{figure}

\emph{Scaling of MFPT to a node with the smallest degree.}  We now study the MFPT in uncorrelated  scale-free networks when the target is a node with the smallest degree.
According to equation~(\ref{C11}), the MFPT to a smallest node can be represented as
\begin{equation}\label{X1}
T_{\rm S}=\frac{\sum_{i=1}^{N}k_i^2}{k_{\rm min}^2}\,,
\end{equation}
where $k_{\rm min}$ denotes the degree of a node with the least degree.
For sparse scale-free networks, their average node degree is a small constant~\cite{Ne03}. Hence the minimal degree $k_{\rm min}$ can be regarded as a smaller constant. Then, recalling equations (\ref{E2}) and (\ref{E3}), the dominant scaling of $T_{\rm S}$ can be approximated by
\begin{eqnarray}\label{X2}
T_{\rm S}\sim\left\{\begin{aligned}
&N^{2/(\gamma-1)}, \quad 2<\gamma<3, \\
&N\ln N, \quad \gamma=3, \\
&N, \quad \gamma>3, \\
\end{aligned}
\right.
\end{eqnarray}
which is supported by extensive numerical simulations, see Fig.~\ref{T-S}.

Equation~(\ref{X2}) indicates that the MFPT $T_{\rm S}$ for MERW in uncorrelated scale-free networks also exhibits rich behavior relying on the degree exponent $\gamma$. When $2<\gamma<3$, $T_{\rm S}$ varies superlinearly with the network size $N$; when $\gamma=3$, $T_{\rm S}$ scales with $N$ as $N\ln N$; when $\gamma>3$, $T_{\rm S}$ behaves linearly with $N$.

Although for both cases that the target is either at a hub node or at a smallest node, the MFPT is influenced by the degree parameter $\gamma$, the dependence relation of MFPT on $\gamma$ is  quite distinct, as can be seen from equations~(\ref{E4}) and~(\ref{X2}). Furthermore, in the whole range of $2<\gamma< \infty$, $T_{\rm H}$ is much smaller than $T_{\rm S}$. As a result, for MERW in uncorrelated scale-free networks, generating all paths with identical probability is disadvantageous for the walker to explore nodes with small degree.


In addition, the MFPT $T_{\rm S}$ for MERW is also different from that of TURW,
where the MFPT $T_{\rm S}$ to a node with the smallest degree can be estimated as
\begin{equation}\label{F2}
T_{\rm S}=\frac{\sum_{i=1}^N k_i}{k_{\rm min}}\approx\sum_{i=1}^N k_i\sim N,
\end{equation}
where equation~(\ref{E6}) is used. This analytical solution is conformed by numerical results shown Fig.~\ref{T-S}.

Equations~(\ref{F2}) implies that for TURW, the MFPT $T_{\rm S}$ scales linearly with network size $N$, independent of $\gamma$, which is totally different from the behavior of MFPT, provided by equation~(\ref{X2}), corresponding to MERW.
From equations (\ref{X2}) and (\ref{F2}), we can observe that for $2<\gamma\leq3$, the MFPT $T_{\rm S}$ for MERW is larger than that for TURW; and that for $\gamma>3$, although the leading scaling of MFPT $T_{\rm S}$ grows linearly with network $N$ for both MERW and TURW, the values of $T_{\rm S}$ for MERW is greater than those corresponding to TURW, which can be seen from  Fig.~\ref{T-S}. Thus, for the case of target node located at a node with the smallest degree, performing MERW is substantially slowly to arrive at the destination than performing TURW, which is in stark contrast with the the case when the target node is a hub node, for which, performing MERW is more efficient than TURW to find the target. This phenomenon can also be accounted for by local transition probability from a node to one of its neighboring nodes with small degree, which is relatively smaller for MERW than for TURW.



\emph{Scaling of MFPT to a uniformly chosen node.} The above studied MFPT to a particular target node in a network is often not looked upon as a general dynamical property of the network~\cite{TeBeVo09,AgBuMa10}. Instead, the average of MFPT over all targets reflects some global characteristics such as the efficiency of searching process. Thus, it is of significance to compute the target averaged MFPT.  Next, we address random walks in an uncorrelated scale-free network with the target being a uniformly selected node in the network. In such situation, the MFPT $\langle T \rangle$ is defined as the average of FPTs over all pairs of nodes in the network, which involves a double average: The former is over all the source nodes to a given target node, the latter is the average of the first one. That is,
\begin{equation}\label{G1}
\langle T\rangle=\frac{1}{N}\sum_{j=1}^N T_j\,.
\end{equation}
Next, we determine $\langle T \rangle$ for MERW and TURW, respectively.

\begin{figure}
\begin{center}
\includegraphics[width=1.1 \linewidth,trim=50 80 0 50]{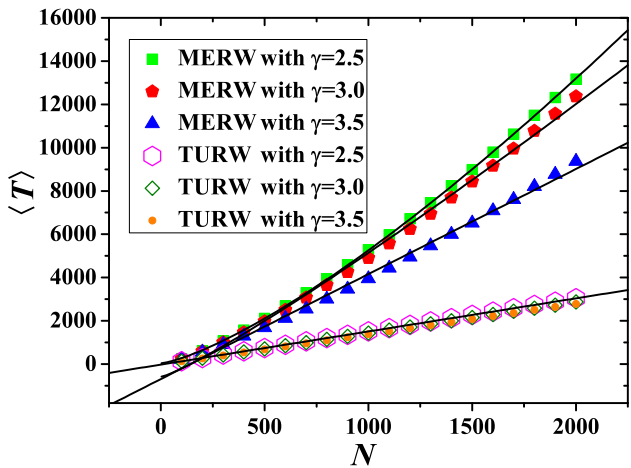}
\end{center}
\caption[kurzform]{\textbf{MFPT to a target node uniformly selected from the whole network.} The filled symbols are the  numerical results generated by equation (\ref{B9}); while the lines correspond to the theoretical predictions given by equations (\ref{X5}) or (\ref{G4}). 
}\label{GATT}
\end{figure}

For MERW, plugging equation~(\ref{C11}) into equation~(\ref{G1}) gives
\begin{equation}\label{X3}
\langle T\rangle=\frac{1}{N}\sum_{j=1}^{N}\frac{1}{k_j^2}\sum_{i=1}^{N}k_i^2\,,
\end{equation}
where the term $\sum_{j=1}^{N}k_j^{-2}$ can be estimated by
\begin{equation}\label{X4}
\sum_{j=1}^{N}\frac{1}{k_j^2}\approx\int_{k_{\rm min}}^{k_{\rm max}}N P(k) k^{-2}dk\sim N\,.
\end{equation}
Considering equations (\ref{E2}), (\ref{E3}) and (\ref{X4}), the quantity $\langle T\rangle$ follows
\begin{eqnarray}\label{X5}
\langle T\rangle\sim\left\{\begin{aligned}
&N^{2/(\gamma-1)}, \quad 2<\gamma<3, \\
&N\ln N, \quad \gamma=3, \\
&N, \quad \gamma>3, \\
\end{aligned}
\right.
\end{eqnarray}
which is consistent with the numerical results, see Fig.~\ref{GATT}.

By comparing equation~(\ref{X2}) and (\ref{X5}), we observe that for MERW the MFPT $\langle T\rangle$ exhibits similar behavior as that of $T_{\rm S}$. This phenomenon
can be heuristically understood from the structure of scale-free networks, where the fraction
of nodes with small degrees is very high. Moreover, the MFPT to a small-degree node is much larger than that of large-degree node. Thus, $\langle T \rangle$ and $T_{\rm S}$ resemble in the leading behavior, which means that the dominant scaling of $T_{\rm S}$ to a
small-degree node is representative of MERW in scale-free networks.


We proceed to uncover the difference for $\langle T\rangle$ between MERW and TURW. For TURW, the quantity $\langle T\rangle$ can be approximated by
\begin{equation}\label{G2}
\langle T\rangle=\frac{1}{N}\sum_{j=1}^N\frac{1}{k_j}\sum_{i=1}^N k_i\,,
\end{equation}
where the term $\sum_{j=1}^N k_j^{-1}$ can be estimated as
\begin{equation}\label{G3}
\sum_{j=1}^N k_j^{-1}\approx\int_{k_{\rm min}}^{k_{\rm max}}N P(k)k^{-1}dk\sim N\,.
\end{equation}
Recalling equation~(\ref{E6}), we have
\begin{equation}\label{G4}
\langle T\rangle\sim N\,,
\end{equation}
a scaling similar to that of $T_{\rm S}$ for TURW. In Fig.~\ref{GATT}, we plot the numerical results of $\langle T\rangle$ versus theoretical prediction in equation~(\ref{G4}) for TURW in scale-free networks with different $\gamma$, which are consistent with each other.


On the other hand, for uncorrelated networks, the relation $\sum_{j=1}^N 1/T_j=1$ holds. Then, according to inequality of arithmetic and geometric means, we can deduce a lower bound of $\langle T\rangle$ for uncorrelated networks:
\begin{equation}\label{G5}
\langle T\rangle=\frac{1}{N}\sum_{j=1}^N T_j\geq\frac{N}{\sum_{j=1}^N 1/T_j}=N\,,
\end{equation}
which provides a minimal scaling for $\langle T\rangle$ in uncorrelated networks.

Equations (\ref{X5}) and (\ref{G4}) show that although the scaling of $\langle T\rangle$ for MERW and TURW behaves differently, the optimal linear scaling for $\langle T\rangle$ can be achieved both for TURW in scale-free networks with arbitrary $2<\gamma<\infty$ and for MERW in scale-free networks with $\gamma>3$. In the end, we stress that although in the case of $\gamma>3$, for both MERW and TURW, $\langle T\rangle$ can reach the minimal scaling, the cofactor of the dominating scaling for $\langle T \rangle$ is larger for MERW than for TURW, which can be seen in Fig.~\ref{GATT}. Therefore, in the whole range of $2<\gamma<\infty$, the value of $\langle T\rangle$ is higher for MERW than that for TURW.




\section*{\large{Discussion}}


In summary, we have presented a comprehensive  and systematical analysis of MFPT for MERW in complex networks. We have provided an explicit expression of MFPT for MERW in a general network with a target node being located at an arbitrary node, which is provided in terms of eigenvalues and eigenvectors of the adjacency matrix for the network. Moreover, for
MERW in an uncorrelated network, we have given an alternative theoretical prediction for MFPT at the mean-field level, which is devoid of calculating the eigenvalues and eigenvectors but gives good approximation for MFPT that are confirmed by extensive numerical results.

Applying the mean-field approximation formula, we have further addressed the leading behavior of MFPT for MERW in uncorrelated scale-free networks with a given target and various degree exponent $\gamma$, focusing on three representative cases with the target being a hub node, or a node with the least links, or a node chosen uniformly. For all the three cases, the MFPT is dependent on  the degree of the target, as well as the degree exponent $\gamma$. We have also performed a comparison of the obtained results for MERW with those corresponding to TURW. For the case that the target is located at a hub node, a walker performing MERW arrives at the destination more quickly than performing TURW. However, for the two cases that target node is a node with the smallest degree or a node selected uniformly, MERW is less efficient for finding the target than for TURW.

We have also found that the values of MFPT for MERW in scale-free networks are distributed over a larger range than their counterparts for TURW. Thus, as an indicator of node importance, MFPT for MERW is better at discriminating influential nodes from common noncentral nodes. Finally, we note that our approximate analytical results only hold for uncorrelated networks. Since in real networks, there exist ubiquitous degree correlations among nodes~\cite{Ne03}, it would be interesting to extend our methods to correlated networks in the future.





\section*{\large{Methods}}

\textbf{Expressing FPT for MERW in a network in terms of the spectra of its adjacency matrix.} It has been reported~\cite{ZhShCh13} that for generic random walks in a weighted network, the FPT $T_{ij}$ from node $i$ to node $j$ can be represented by the eigenvalues and eigenvectors of the following matrix $\bf\Gamma$ defined as
\begin{equation}\label{B1}
{\bf\Gamma}={\bf S}^{-\frac{1}{2}}{\bf W}{\bf S}^{\-\frac{1}{2}}={\bf S}^{\frac{1}{2}}{\bf P}{\bf S}^{-\frac{1}{2}},
\end{equation}
where $\bf S$ is the diagonal strength matrix with its $i$th diagonal entry equal to the strength $s_i$ of node $i$. It is evident that matrix $\bf\Gamma$ is real and similar to the transition matrix $\bf P$ and thus has the same set of eigenvalues as $\bf P$.
Let $\lambda_1^{\rm P}, \lambda_2^{\rm P}, \cdots, \lambda_3^{\rm P}$ be the $N$ eigenvalues of matrix $\bf\Gamma$ for a network of size $N$, rearranged as $1=\lambda_1^{\rm P}>\lambda_2^{\rm P}\geq\cdots\geq\lambda_N^{\rm P}$, and let $\psi_1, \psi_2, \cdots, \psi_N$ denote the corresponding normalized and mutually orthogonal eigenvectors, where $\psi_i=(\psi_{i1}, \psi_{i2}, \cdots, \psi_{iN})^{\top}$. Then, the FPT for a walker starting from node $i$ to first arrive at node $j$ can be expressed as~\cite{ZhShCh13}
\begin{equation}\label{B2}
T_{ij}=\frac{s}{s_j}\sum_{k=2}^N\frac{1}{1-\lambda_k^{\rm P}}\left(\psi_{kj}^2-\psi_{ki}\psi_{kj}\sqrt{\frac{s_j}{s_i}}\right).
\end{equation}

For the particular weighted network associated with MERW, its weight matrix $\bf W$ satisfies
\begin{equation}\label{B3}
{\bf W}=\frac{1}{\lambda_1}{\bf S}^{\frac{1}{2}}{\bf A}{\bf S}^{\frac{1}{2}}.
\end{equation}
Inserting equation~(\ref{B3}) into equation~(\ref{B1}) leads to
\begin{equation}\label{B4}
{\bf \Gamma}=\frac{1}{\lambda_1}{\bf A}\,,
\end{equation}
which implies that the eigenvalues and eigenvectors of the two matrices $\bf\Gamma$ and $\bf A$, satisfy the following one-to-one relations:
\begin{equation}\label{B5}
\lambda_i^{\rm P}=\frac{\lambda_i}{\lambda_1}
\end{equation}
and
\begin{equation}\label{B6}
\psi_i=\mu_i.
\end{equation}
Substituting equations~(\ref{B5}) and (\ref{B6}) into equation~(\ref{B2}) and considering $s_i=\lambda_1\mu_{1i}^2$, we obtain
\begin{equation}\label{B7}
T_{ij}=\frac{1}{\mu_{1j}^2}\sum_{k=2}^N\frac{\lambda_1}{\lambda_1-\lambda_k}\left(\mu_{kj}^2-\mu_{ki}\mu_{kj}\frac{\mu_{1j}}{\mu_{1i}}\right)\,,
\end{equation}
which provides a close-form expression of the FPT for MERW starting from an arbitrary node $i$ to another node $j$. 


\textbf{Approximation of the principal eigenvector.} Mean-field theory assumes that nodes having the same degree share the same structural properties~\cite{BaPa10}. Based on this hypothesis, we use $\mu(k)$ to denote the value of elements of principal eigenvector corresponding to nodes with degree $k$. By definition,
\begin{equation}\label{C3}
{\bf A}\mu_1=\lambda_1\mu_1\,.
\end{equation}
Applying the coarse-graining idea to degree classes~\cite{BoCaPa09}, equation~(\ref{C3}) is equivalent to
\begin{equation}\label{C4}
{\bf\bar A}\bar\mu=\lambda_1\bar\mu\,,
\end{equation}
where $\bar\mu=(\mu(k_1), \mu(k_2), \cdots, \mu(k_N))^\top$. The entry $\bar a_{k_i k_j}$ of matrix $\bf\bar A$ defines the probability that two nodes of degree $k_i$ and $k_j$ are adjacent, namely
\begin{equation}\label{C5}
\bar a_{k_i k_j}=\frac{k_i P(k_j|k_i)}{N P(k_j)},
\end{equation}
where $P(k_j|k_i)$ is the conditional probability~\cite{PaVaVe01} that a node of degree $k_i$ is directly connected to a node with degree $k_j$. Note that $\bf\bar A$ can be also interpreted as a weight matrix of a weighted, fully connected graph, which is obtained by annealed network approach~\cite{DoGoMe08}. Since $\bar\mu$ is the eigenvector of $\bf \bar A$ corresponding to the principle eigenvalue $\lambda_1$, we can approximate $\mu_1$ by $\bar\mu$. Next, we  evaluate the principle eigenvector $\bar\mu$ of $\bf\bar A$.

Since in an uncorrelated network, the degrees of the two nodes connecting any edge are completely independent, the conditional probability can be estimated as
\begin{equation}\label{C6}
P(k_j|k_i)=k_j P(k_j)/\langle d\rangle\,,
\end{equation}
where $\langle d\rangle$ is the average node degree. Instituting equation~(\ref{C6}) into equation~(\ref{C5}) yields
\begin{equation}\label{C7}
\bar a_{k_i k_j}=\frac{k_i k_j}{N\langle d\rangle},
\end{equation}
which indicates that the matrix $\bf\bar A$ can be represented as
\begin{equation}\label{C8}
{\bf\bar A}=\frac{1}{N\langle d\rangle}{\bf k}{\bf k}^{\top}\,,
\end{equation}
where ${\bf k}$ is the degree sequence of the network and can be denoted by a vector as
\begin{equation}\label{C9}
{\bf k}=(k_1, k_2, \cdots, k_N)^{\top}\,.
\end{equation}
For a matrix having the form ${\bf\alpha}{\bf \alpha}^{\top}$, where $\bf\alpha$ is a nonzero vector, its rank is $1$. Therefore, the rank of $\bf\bar A$ is $1$, and $\bf\bar A$ has exactly one nonzero eigenvalue
\begin{equation}\label{D2}
\lambda={\rm tr}({\bf \bar A})=\frac{1}{N\langle d\rangle}\sum_{i=1}^{N}k_i^2\,.
\end{equation}

Having obtained the principal eigenvalue $\lambda$, we continue to determine its corresponding eigenvector $\bar\mu$. Obviously, $\lambda$ and  $\bar\mu$ satisfy the following relation:
\begin{equation}\label{D3}
{\bf \bar A}\bar\mu=\lambda\bar\mu\,,
\end{equation}
which can be reexpressed as a system of equations:
\begin{eqnarray}\label{D4}
\left\{\begin{aligned}
k_1\bar\mu_1+k_2\bar\mu_2+\cdots+k_N\bar\mu_N&=\frac{\lambda}{k_1}\bar\mu_1,\\
k_1\bar\mu_1+k_2\bar\mu_2+\cdots+k_N\bar\mu_N&=\frac{\lambda}{k_2}\bar\mu_2,\\
&\vdots\\
k_1\bar\mu_1+k_2\bar\mu_2+\cdots+k_N\bar\mu_N&=\frac{\lambda}{k_N}\bar\mu_N,\\
\end{aligned}
\right.
\end{eqnarray}
where $\bar\mu_i$ is the $i$th entry of $\bar\mu$. Therefore,
\begin{equation}\label{D5}
\frac{\bar\mu_1}{k_1}=\frac{\bar\mu_2}{k_2}=\cdots=\frac{\bar\mu_N}{k_N}.
\end{equation}
Combining with normalized condition $\sum_{i=1}^{N}\bar\mu_i^2=1$, equation~(\ref{D5}) can be solved to obtain
\begin{equation}\label{D6}
\bar\mu_j^2=\frac{k_j^2}{\sum_{i=1}^{N}k_i^2}\,.
\end{equation}
Approximate $\mu_{1j}$ by $\bar{\mu}_j$ leads to
\begin{equation}\label{Z4}
\mu_{1j}^2\approx \frac{k_j^2}{\sum_{i=1}^N k_i^2}\,.
\end{equation}

\section*{\large{Acknowledgments}}
This work was supported by the National Natural Science Foundation of China under Grants No. 11275049 and the National Basic Research Program of China under Grant No. 2010CB731401.

\section*{\large{Author contributions}}
Y.L. and Z.Z.Z. designed the research, performed the
research, and wrote the manuscript.

\section*{\large{Additional information}}

\textbf{Competing financial interests:} The authors declare no competing financial interests.

\end{document}